\documentclass[aps,physrev,reprint,superscriptaddress]{revtex4-2}
\usepackage{graphicx}% Include figure files
\graphicspath{{../}}
\usepackage{caption} % 用于图注格式设置
\usepackage{hyperref}
\usepackage[T1]{fontenc}

\usepackage{bm}% bold math
\usepackage{ragged2e}  % 关键宏包：提供 \justify 命令，支持两端对齐
\usepackage[version=4]{mhchem}

\begin{document}

%\preprint{}

\title{Distinct routes to phase transitions in spatial activation systems}
%Distinct routes to collective activation in spatial threshold systems
%Interaction range governs the routes to collective activation in spatial threshold systems
%Characteristic Interaction Range Governs the Crossover between Gradual Expansion, Nucleation, and Branching in Spatial Bootstrap Percolation


%\thanks{}

\author{Jialu Zhang}
\affiliation{Department of Systems Science, Faculty of Arts and Sciences, Beijing Normal University, Zhuhai 519087, China}
\affiliation{International Academic Center of Complex Systems, Beijing Normal University, Zhuhai, 519087, China}
\affiliation{School of Systems Science, Beijing Normal University, Beijing, 100875, China}

\author{Guanyu Zhang}
\affiliation{Department of Systems Science, Faculty of Arts and Sciences, Beijing Normal University, Zhuhai 519087, China}
\affiliation{International Academic Center of Complex Systems, Beijing Normal University, Zhuhai, 519087, China}

\author{Leyang Xue}
\email{hsuehleyang@gmail.com}
\affiliation{Department of Physics, Bar-Ilan University, Ramat-Gan, 52900, Israel}

\author{Peng-Bi Cui}
\email{cuisir610@gmail.com}
\affiliation{Department of Systems Science, Faculty of Arts and Sciences, Beijing Normal University, Zhuhai 519087, China}
\affiliation{International Academic Center of Complex Systems, Beijing Normal University, Zhuhai, 519087, China}
\affiliation{School of Systems Science, Beijing Normal University, Beijing, 100875, China}


\date{\today}

\begin{abstract}
Threshold-driven activation governs a wide range of collective phenomena, yet the microscopic origins of its phase transitions in spatial systems remain unresolved.
Here, we show that spatial activation systems undergo multiple distinct routes to phase transitions, controlled by a single parameter---the interaction range.
We uncover a unified phase diagram featuring continuous, first-order, and mixed-order transitions, and demonstrate that the two abrupt transitions arise from fundamentally different mechanisms: nucleation-driven front propagation and critical branching.
These routes exhibit distinct dynamical scaling, establishing a direct link between microscopic activation dynamics and macroscopic critical behavior.
We further identify a metastable phase in which global activation cannot be achieved by random activation alone, but can be triggered by localized seeds.
In this regime, the critical activation nucleus remains finite and independent of system size, implying that arbitrarily large systems can remain stable under random perturbations yet highly vulnerable to localized triggers.
The onset of this phase is abrupt, revealing an extreme sensitivity of collective dynamics to small parameter changes.
These results establish a mechanistic framework for phase transitions in spatial activation systems and reveal how microscopic perturbations can trigger macroscopic cascades.

\end{abstract}

%\keywords{}

\maketitle 
 
\section{\label{sec:introduction}Introduction}
Threshold-driven activation process is ubiquitous in complex systems across physical~\cite{bak1987soc,olami1992ofc,sethna2001crackling}, biological~\cite{amini2010bootstrap,goltsev2010stochastic}, and social domains~\cite{centola2010spread,daqing2014spatial,zhao2016spatio}.
In many cases, the functional state of each unit can be coarse-grained into a binary variable (inactive or active).
Activation occurs only when the cumulative influence from its neighborhood exceeds a threshold, allowing microscopic interactions to produce macroscopic collective behavior.
The resulting states exhibit a rich spectrum of phase transitions, ranging from localized to global activation, often governed by the underlying cascade mechanism~\cite{holroyd2003sharp,baxter2010bootstrap,gao2015bootstrap}.
Understanding these mechanisms is therefore essential for revealing and ultimately controlling collective activation in complex systems.

Current understanding of threshold-activated systems largely stems from non-spatial settings, where interaction ranges between units are unconstrained by geometry~\cite{chalupa1979bootstrap,baxter2010bootstrap}.
In this regime, local perturbations can influence distant parts of the network and trigger system-wide cascades, giving rise to mixed-order phase transitions~\cite{wu2014multiple}.
However, this picture changes fundamentally in spatially embedded systems.
Here, interactions are predominantly local, generating spatial correlations and modifying the effective contact structure.
As a result, spatial proximity can qualitatively reshape cascade mechanisms and the resulting macroscopic dynamics.
For example, in epidemics, distance-limited contacts can slow the spread and alter how outbreaks persist or re-merge~\cite{firestone2011importance,zhang2016modeling}.
In neuronal systems, spatial constraints introduce wiring costs that shape the propagation of neural activity~\cite{roberts2016contribution,pang2023geometric,zhang2024geometric}.
In traffic networks, cascades are primarily driven by local congestion rather than long-range interactions~\cite{daqing2014spatial,li2015percolation,li2019vulnerability}.

Empirical evidence across domains further indicates that interaction range can qualitatively reshape the dynamics of spatial threshold systems.
For example, in bacterial colonies, nutrient availability effectively tunes interaction range, shifting growth from ramified (branched) patterns to nucleation-driven compact expansion~\cite{fujikawa1989fractal,ben1994generic}.
In epidemic spreading, increased mobility expands the effective interaction range, transforming propagation from coherent traveling waves to rapid, spatially fragmented outbreaks~\cite{grenfell2001travelling,hallatschek2014acceleration}.
In neural tissue, local coupling supports wave-like $\mathrm{Ca}^{2+}$ activity, whereas longer-range connectivity enables abrupt, system-wide synchronization~\cite{lechleiter1991spiral}.
These observations suggest that interaction range acts as a key control parameter, setting the spatial scale over which local threshold events accumulate into global activation~\cite{niemeyer1984fractal,breskin2006percolation,brockmann2013hidden}.

Despite their ubiquity, the mechanisms by which interaction range shapes collective activation remain poorly understood.
In many studies, interaction range is treated as a context-dependent feature and is intertwined with system-specific interaction rules~\cite{danziger2013interdependent,gao2015bootstrap,zhang2025delayed}.
This makes it difficult to isolate its dynamical role and to compare different spatial systems on equal footing.
Here, we address this gap using bootstrap percolation (BP) as a minimal model of threshold activation~\cite{chalupa1979bootstrap,baxter2010bootstrap}.
In BP, a node becomes permanently active once the number of its active neighbors exceeds a prescribed threshold.
To incorporate spatial interaction range, we embed BP on spatial networks generated by the $\zeta$-model~\cite{danziger2016effect,xue2024nucleation}.
Nodes are placed on a two-dimensional square lattice with periodic boundary conditions.
At fixed mean degree $\langle k\rangle$, links are added randomly with Euclidean lengths $l$ drawn from an exponential distribution, $P(l)\sim e^{-l/\zeta}$.
By tuning $\zeta$, we isolate the effect of interaction range while avoiding additional microscopic assumptions.

%To our best knowledge, the prevailing paradigm regards effective interaction range as one requisite parameter for consideration, and routinely integrate it with the specific nature of inter-unit interactions in diverse spatial threshold-driven complex systems to elucidate the dynamics of their own emergent features~\cite{chalupa1979bootstrap,baxter2010bootstrap,peng2025bootstrap,cerf1999finite,holroyd2003sharp,danziger2013interdependent,gao2015bootstrap,koch2016bootstrap,zhang2025delayed}. This obscures the general and essential dynamical effects of this parameter. We are thus unable to definitively answer whether there exists a universal dynamical principle that underlies the emergence routes. To address this challenge, it necessitates a minimal framework anchored in this fundamental control parameter.

%In this work, we investigate bootstrap percolation on spatially embedded two-dimensional networks with a tunable interaction range, $\zeta$.
%Beyond mean-field descriptions~\cite{saberi2015recent, wu2014multiple}, we show that interaction range acts as an organizing principle for collective activation, selecting how activation unfolds in space and time.

%With this setup, we now turn to the key findings of our study.
Our results show that, as the initial activation probability $p$ increases, the system exhibits a double transition.
The lower critical point, $p_c^{(1)}$, corresponds to a connectivity transition: it is continuous and largely independent of the interaction range $\zeta$.
We focus on the higher critical point, $p_c^{(2)}$, where genuine collective activation emerges.
The nature of the transition at $p_c^{(2)}$ depends strongly on $\zeta$ and reflects distinct microscopic cascade mechanisms.
For short interaction ranges ($\zeta < \zeta_c$), activation spreads locally through fractal growth, yielding a continuous transition.
For intermediate ranges ($\zeta \gtrsim \zeta_c$), activation is nucleation-driven, leading to a purely first-order transition.
When long-range connections become comparable to the system size ($\zeta \sim L$), activation is dominated by system-wide branching, and the transition becomes mixed-order, featuring an abrupt jump accompanied by critical scaling, as observed in random networks~\cite{baxter2010bootstrap}.

Beyond these transition behaviors, our numerical and theoretical results further reveal three distinct dynamical phases in spatial systems: (i) an activatable phase, (ii) an inactivatable phase, and (iii) a metastable phase.
In the activatable phase, random perturbations alone can trigger global activation.
In contrast, in the inactivatable phase, the system remains inactive even when locally activated regions grow to nearly the system size.
Between them lies a metastable phase: random seeding typically fails to trigger a global cascade, yet a localized microscopic nucleus can still initiate propagation and eventually activate the entire system.
In this regime, the system is robust to random perturbations but highly sensitive to localized fluctuations, such that controlling only a small fraction of nodes can alter the macroscopic state.
Notably, the boundaries between these phases are sharp and lack reliable early-warning indicators: even small changes in interaction range or initial activation can rapidly switch the system between phases.
This behavior reveals that spatial threshold-driven systems can be extremely sensitive to external interventions.

Our results uncover the mechanisms that govern macroscopic activation and provide a unified explanation for distinct type of phase transitions.
In particular, we demonstrate that abrupt transitions can arise from fundamentally different mechanisms---nucleation and branching, revealing distinct routes to collective activation in spatial systems.
These findings show that, in this metastable regime, macroscopic activation can be controlled through localized perturbations.
%^Together, within a single modeling framework, we uncover three distinct routes to different transition behaviors, underlie which there are three different activation mechanisms, as evidenced by their scaling behaviors. Indeed, in any case, interaction range determines how a system responds to perturbations, whether activation persists, and the manner and sensitivity with which it unfolds.
% More broadly, our results show that collective activation in spatial threshold systems cannot be understood solely in terms of connectivity, activation thresholds, or initial conditions.
% Rather, interaction range determines how a system responds to perturbations, shaping not only whether activation occurs, but also the manner and sensitivity with which it unfolds.
% By establishing interaction range as a fundamental control parameter linking spatial constraints, activation mechanisms, metastability, and cascading dynamics, this work provides new insight into understanding---and potentially controlling---the behavior of real-world spatial threshold-driven systems.

\section{\label{sec:model}Model}
We study classical bootstrap percolation (BP) on a spatial network with a tunable interaction range.
The network is constructed using the $\zeta$-model~\cite{danziger2016effect,xue2024nucleation}: $N=L\times L$ nodes are placed on a two-dimensional periodic square lattice, and link lengths $l$ are drawn from $P(l)\propto e^{-l/\zeta}$, with links assigned randomly to satisfy the prescribed mean degree $\langle k \rangle$.
The parameter $\zeta$ sets the characteristic interaction range and provides a continuous interpolation between spatially constrained and effectively non-spatial topologies.
When $\zeta \ll L$, long links are rare, and the network is close to a two-dimensional lattice with only small degree fluctuations.
As $\zeta \to L$, link lengths become effectively independent of geometric distance, and the system approaches a non-spatial random network.
A representative realization of the network is shown in Fig.~\ref{fig:1}(a).
Details of the implementation are provided in Appendix~\ref{sec:net_model}.

Bootstrap percolation on this network begins with random seeding: each node $i$ is active with probability $p$ and inactive with probability $1-p$.
The system then evolves iteratively under the standard BP rule: an inactive node becomes active once at least $T$ of its neighbors are active.
The process continues until no further activations occur, yielding a stable configuration of active clusters.
In many systems, the largest active cluster reflects the system's functional state; accordingly, we use its size as the order parameter throughout this work.
Figure~\ref{fig:1}(a) illustrates this process on a small network ($T=2$).
Unless noted otherwise, all results in the main text use $T=6$; results for other thresholds are given in the Supplementary Information.

% To characterize the macroscopic outcome of the activation process, we use the relative size of the giant connected component (GCC) of the final activation cluster, denoted by $P_\infty$, as the order parameter.
% As the initial activation probability $p$ is increased incrementally, two distinct critical points can be identified.
% The first critical point, $p_c^{(1)}$, corresponds to the emergence of a giant connected component in the final activation cluster, at which $P_\infty$ first becomes nonzero.
% This transition reflects a purely structural change: the initially activated seeds are sufficient to form a macroscopic connected cluster, but activation remains localized and does not yet propagate across the system.
% The second critical point, $p_c^{(2)}$, marks a genuine dynamical transition from localized activation to system-wide propagation, where activation spreads across a finite fraction of the system.
% This higher threshold signals the onset of collective activation and constitutes the primary focus of this work.

\begin{figure*}[htbp]
  \centering
  % 导入包含a-d子图的主图
  \includegraphics[width=\textwidth]{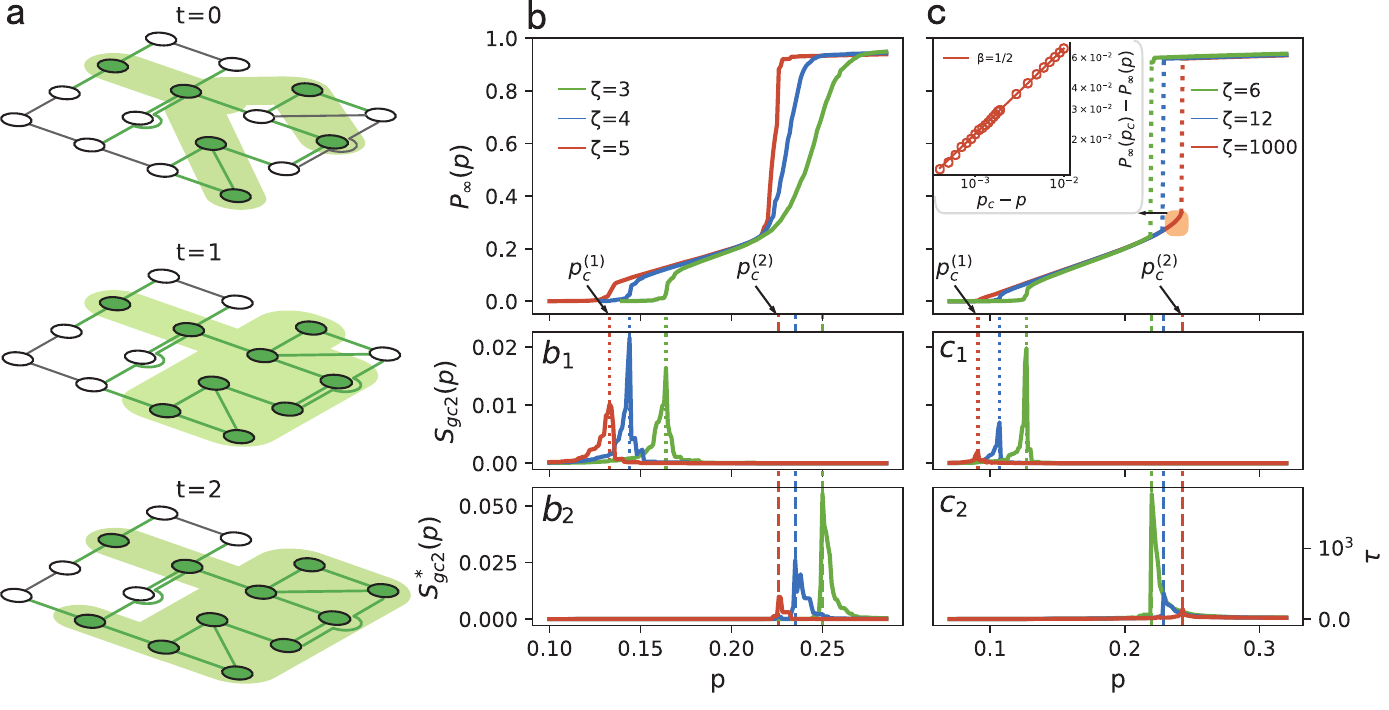}

  % 定义子图标签（不显示独立子图，仅用于编号和引用）
  %\begin{subcaptiongroup}  % 统一管理子图编号
  %    \phantomsubcaption\label{fig:1a}  % 子图a标签
  %    \phantomsubcaption\label{fig:1b}  % 子图b标签
  %    \phantomsubcaption\label{fig:1c}  % 子图c标签
  %\end{subcaptiongroup}

  % 主图图注：明确(a)-(d)对应主图内的区域
  \caption{\textbf{Bootstrap percolation on spatial networks reveals interaction-range--dependent activation transitions.}
    \textbf{(a)} Illustration of an irreversible bootstrap percolation process with threshold $T=2$ on a two-dimensional spatial network generated by the $\zeta$-model ($N=16$, $\langle k \rangle=3$, $\zeta=1$).
    Initially active nodes are shown in green at $t=0$.
    An inactive node becomes active once at least $T$ of its neighbors are active.
    The cascade terminates at $t=2$, when no further activations are possible.
    \textbf{(b,c)} Bootstrap percolation with threshold $T=6$ for different interaction ranges $\zeta$.
    Panels (b,c) show the relative size of the giant active cluster, $P_\infty$, as a function of the initial activation probability $p$.
    In all cases, a \emph{double phase transition} is observed:
    a lower transition at $p_c^{(1)}$ marking the emergence of a macroscopic connected cluster, and a higher transition at $p_c^{(2)}$ corresponding to the onset of genuine collective activation.
    Importantly, interaction range qualitatively reshapes the nature of the second transition $p_c^{(2)}$.
    While the transition is continuous at small $\zeta$, it becomes discontinuous as $\zeta$ increases.
    Yet discontinuity alone does not imply a unique mechanism.
    As revealed by the inset, the transition at $\zeta=1000$ remains discontinuous but exhibits critical scaling.
    \emph{Inset:} Near $p_c^{(2)}$ for $\zeta=1000$, $P_\infty$ exhibits critical scaling $|P_\infty(p_c^{(2)})-P_\infty(p)| \sim (p_c^{(2)}-p)^\beta$ with $\beta=1/2$, a hallmark of mixed-order behavior.
    (\textbf{$b_1$,$c_1$}) Relative size of the second-largest active component, $S_{gc2}$, as a function of $p$.
    The peak of $S_{gc2}(p)$ identifies the lower critical point $p_c^{(1)}$.
    (\textbf{$b_2$}) Relative size of the second-largest component composed of inactive nodes, $S^*_{gc2}$, versus $p$, whose maximum identifies the continuous critical point at $p_c^{(2)}$.
    (\textbf{$c_2$}) Number of cascade iterations $\tau$ required to reach a stable state as a function of $p$.
    A sharp peak in $\tau$ signals an abrupt transition at $p_c^{(2)}$.
    All results are obtained for networks of size $N=1000\times1000$ with mean degree $\langle k \rangle=10$.
    For continuous transitions, data are averaged over 10 independent realizations, while a single representative realization is shown for abrupt transitions.
    Results for other threshold values $T$ are reported in the Supplementary Information.
  }
  \label{fig:1}  % 主图标签
\end{figure*}

\section{\label{sec:results}Results}
\subsection{\label{subsec:phase_transition}Phase transition behavior}
We show that the system undergoes a double phase transition, consistent with earlier findings~\cite{gao2015bootstrap}; 
More importantly, the interaction range controls the nature of the transition and governs the emergence of distinct cascade patterns.
For each fixed $\zeta$ (Figs.~\ref{fig:1}(b) and (c)), increasing $p$ reveals two distinct critical points, $p_c^{(1)}$ and $p_c^{(2)}$, corresponding to different dynamical roles in the cascade process.
The first critical point, $p_c^{(1)}$, is located by the peak of the second-largest active cluster $S_{gc2}$ (Fig.~\ref{fig:1}($b_1$) and ($c_1$)). 
This point corresponds to a structural connectivity transition, where a giant component first emerges.
The second point, $p_c^{(2)}$, is identified by the peak of $S^*_{gc2}$ in the continuous regime (Fig.~\ref{fig:1}($b_2$)) and by the peak of the total iteration number $\tau$ in the discontinuous regime (Fig.~\ref{fig:1}($c_2$)). 
This point marks the onset of a system-wide cascade, where activation becomes self-sustaining and spreads across the network.
%更改：删除下一句话（For different $\zeta$, ...... the interaction range.）
%For different $\zeta$, the lower transition at $p_c^{(1)}$ remains continuous and is close to the classical percolation threshold $p_c$ (see Supplementary Fig.~xxx), whereas the transition nature at $p_c^{(2)}$ depends strongly on the interaction range.

Varying $\zeta$ yields three distinct transition regimes at $p_c^{(2)}$, each governed by a different activation mechanism.
For $\zeta=3,4,5$ (Fig.~\ref{fig:1}(b)), the transition is continuous.
In this short-range regime, long links are too rare to couple distant clusters, so activation remains localized and cannot develop into a system-spanning cascade.
As $\zeta$ increases, long-range links become more prevalent, and the transition at $p_c^{(2)}$ becomes discontinuous (Fig.~\ref{fig:1}(c)).
However, the discontinuous behavior at intermediate ranges ($\zeta=6,12$) differs qualitatively from that in the long-range limit ($\zeta=1000$).
In the long-range limit ($\zeta\sim L$), $P_{\infty}(p)$ shows a clear curvature change near $p_c^{(2)}$, indicating critical behavior.
As shown in the inset of Fig.~\ref{fig:1}(c), near $p_c^{(2)}$ we find $P_{\infty}(p)\sim (p-p_c^{(2)})^{\beta}$ with $\beta=1/2$ \cite{baxter2010bootstrap,saberi2015recent}, consistent with bootstrap percolation on random networks.
The coexistence of critical scaling with a finite jump indicates a mixed-order transition.
By contrast, for intermediate ranges $\zeta \gtrsim \zeta_c \approx 6$, $P_{\infty}$ exhibits an abrupt jump at $p_c^{(2)}$ without observable critical scaling, consistent with a pure first-order transition.
As detailed below, these two discontinuous transitions are driven by distinct activation mechanisms.

\begin{figure*}[htbp]
  \centering
  % 导入包含a-d子图的主图
  \includegraphics[width=\textwidth]{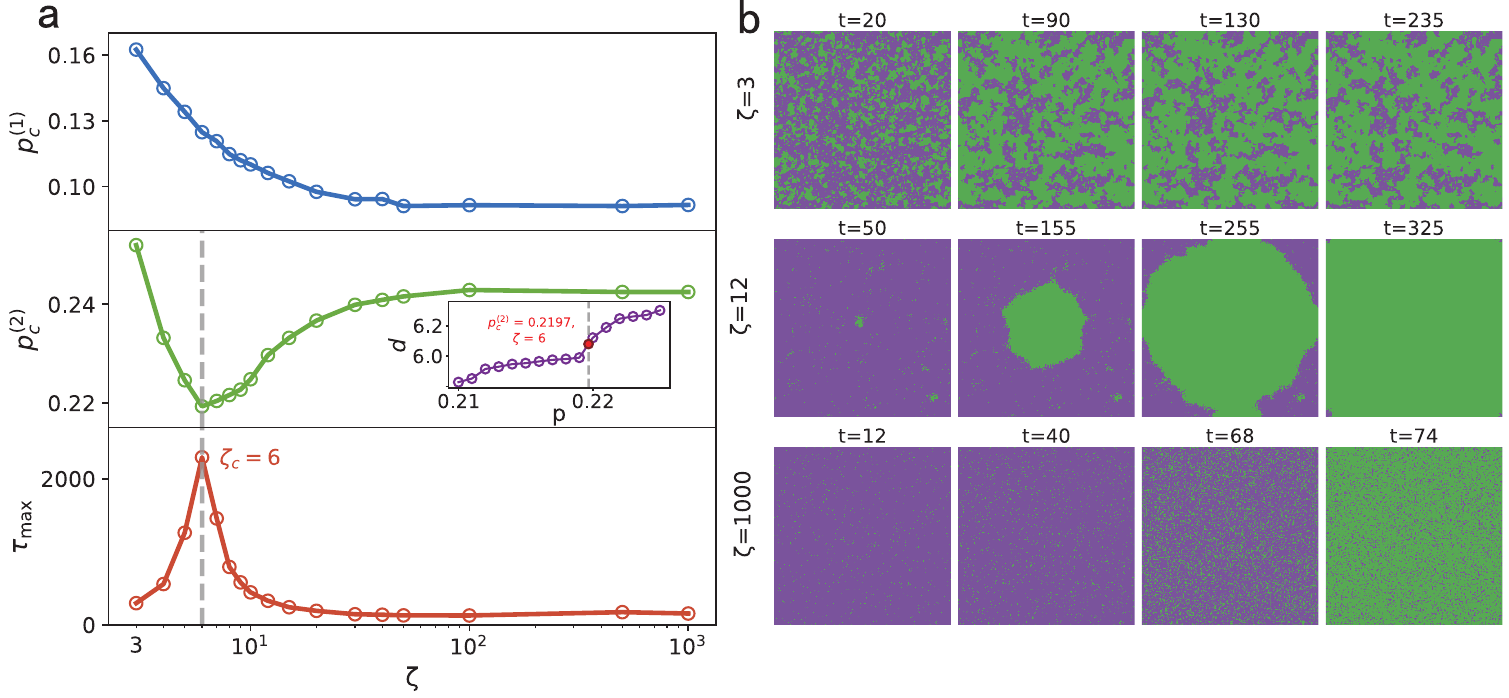}

  % 定义子图标签（不显示独立子图，仅用于编号和引用）
  %\begin{subcaptiongroup}  % 统一管理子图编号
  %    \phantomsubcaption\label{fig:2a}  % 子图a标签
  %    \phantomsubcaption\label{fig:2b}  % 子图b标签
  %\end{subcaptiongroup}
  % 主图图注：
  \caption{\textbf{Interaction range $\zeta$ selects the route and nature of the second phase transition in spatial bootstrap percolation.}
  \textbf{(a)} Upper panel: Critical activation probabilities $p_c^{(1)}$ and $p_c^{(2)}$ as functions of the characteristic interaction range $\zeta$.
    The second critical point $p_c^{(2)}$ exhibits a pronounced minimum at $\zeta_c \simeq 6$ ($p_c^{(2)} \approx 0.2197$), which demarcates two distinct regimes:
    for $\zeta < \zeta_c$, the transition at $p_c^{(2)}$ is continuous;
    for $\zeta \gtrsim \zeta_c$, it becomes discontinuous phase transition;
    %and for $\zeta \sim L$, it crosses over to a mixed-order transition characteristic of non-spatial networks.
    \emph{Inset:} Gyration diameter $d$ of the largest connected cluster during classical percolation on a two-dimensional lattice as a function of the occupation probability $p$.
    When $p = p_c^{(2)}(\zeta_c \simeq 6)$ (dashed line), the gyration diameter reaches $d \simeq 6$, matching the minimal spatial extent required for a nucleus to become self-sustaining and trigger nucleation-driven activation.
    Lower panel: Maximum number of iterations $\tau_{\max}$ required to reach a stable state as a function of $\zeta$.
    The peak of $\tau_{\max}$ at $\zeta_c \simeq 6$ signals critical slowing down associated with the crossover between continuous and discontinuous activation mechanisms.
    \textbf{(b)} Snapshots of the spatial configuration of the giant connected component at several iteration steps $t$ for $p = p_c^{(2)}$, illustrating the distinct activation processes across regimes.
    For small $\zeta$, activation proceeds via fractal, spatially constrained growth, leading to a continuous transition.
    At intermediate $\zeta \gtrsim \zeta_c$, localized nuclei emerge and propagate outward, producing a nucleation-driven, purely first-order transition.
    For $\zeta \sim L$, activation events occur throughout the system and trigger branching-driven avalanches, resulting in a mixed-order transition.
    Green and purple dots represent active and inactive nodes, respectively.
    Iteration steps correspond to those shown in Fig.~\ref{fig:3} for the same values of $\zeta$.
  }
\label{fig:2}  % 主图标签
\end{figure*}

We further examine the microscopic cascade dynamics in each regime.
The snapshots (Fig.~\ref{fig:2}(b)) show that each regime is governed by a distinct mechanism.
For short range ($\zeta=3$, top row), activation forms many small clusters across the system.
These clusters grow locally but rarely merge into a spanning component, as the limited interaction range suppresses long-distance coupling.
Activation thus proceeds through gradual coalescence of localized clusters, yielding a continuous transition with fractal-like spatial organization.
As $\zeta$ increases to the intermediate regime ($\zeta=12$, middle row), the mechanism changes: a markedly larger active cluster emerges.
Once this cluster exceeds the critical nucleation size (around $t=50$), it expands outward in a self-sustained manner and eventually activates the entire system.
This behavior resembles classical nucleation in gas--liquid phase transitions~\cite{debenedetti2020metastable}, where a critical droplet grows and drives a pure first-order transition.
As $\zeta$ approaches the system length ($\zeta\sim L$, $\zeta=1000$, bottom row), spatial locality is effectively lost.
Nodes can then influence distant parts of the network, and activation events become spatially distributed and random.
As these events accumulate, the system reaches a threshold at which a local trigger launches a system-wide avalanche, leading to full activation (between $t=68$ and $74$).
This cascade pathway differs from nucleation-driven growth and is characteristic of a mixed-order transition.
Overall, increasing $\zeta$ reorganizes the microscopic cascade dynamics, producing a crossover from local spreading to nucleation and then to long-range branching.

Distinct cascade mechanisms leave clear signatures in the critical thresholds.
Figure~\ref{fig:2}(a) shows how the two transition points, $p_c^{(1)}$ and $p_c^{(2)}$, vary with the interaction range $\zeta$.
The lower threshold $p_c^{(1)}$ decreases monotonically with $\zeta$, reflecting a geometric effect: increasing the interaction range weakens spatial confinement and enhances connectivity between neighborhoods.
By contrast, $p_c^{(2)}$ is governed by the cascade mechanism and shows a nonmonotonic dependence on $\zeta$.
Starting from the short-range regime, increasing $\zeta$ enhances local cooperative activation among nearby clusters, reducing the cost of sustaining activation and lowering $p_c^{(2)}$.
This trend reaches a minimum at $\zeta_c=6$, where the system is most susceptible to nucleation-driven activation.
Near $p\approx p_c^{(2)}$, the largest activated cluster attains a diameter comparable to the interaction range at $\zeta_c$ (inset of Fig.~\ref{fig:2}).
This indicates that cooperative activation extends across a full interaction neighborhood.
At this scale, spontaneously formed clusters can generate propagating activation fronts.
This corresponds, in the theoretical framework (see Appendix~\ref{sec:theory_rc}), to the onset of a critical self-sustaining activation region.
Nucleation is thus most efficient at $\zeta_c$, giving rise to the minimum of $p_c^{(2)}$.
This picture is supported by the cascade dynamics (Fig.~\ref{fig:2}(a), third row): the maximal cascade time $\tau_{\max}$ exhibits a pronounced peak at $\zeta_c$, indicating critical slowing down at the crossover between activation mechanisms.
Away from this optimal scale, nucleation becomes less efficient.
For $\zeta<\zeta_c$, activation remains too localized to form a self-sustaining front.
For $\zeta>\zeta_c$, long-range links dilute local reinforcement and the dynamics crosses over to branching-like activation.
In both cases, a larger initial active fraction is required, leading to an increase in $p_c^{(2)}$ away from $\zeta_c$.
As $\zeta$ approaches the system size $L$, the cascade mechanism reduces to a branching process in the mean-field regime, and $p_c^{(2)}$ converges to the theoretical value for random networks (i.e., Erd\H{o}s--R\'enyi networks; see Appendix~\ref{sec:theoretical_threshold}).

Beyond the $T=6$ case, we examine how the interaction range $\zeta$ shapes bootstrap percolation across different thresholds $T$.
We find that the macroscopic transition behavior is jointly controlled by $T$ and $\zeta$.
At both low and high $T$, the system exhibits a single transition, driven by rapid activation at low $T$ and by connectivity constraints at high $T$.
At intermediate $T$, a double transition emerges, similar to the $T=6$ case, indicating a separation between connectivity onset and full activation.
In this regime, the effect of $\zeta$ on the transition behavior remains qualitatively unchanged (Supplementary Figs.~S1 and S2).
These results establish $\zeta$ as a key control parameter governing both the nature of the phase transition and the underlying cascade mechanism.

% The lower panel of Fig.~\ref{fig:2}(a) further supports this picture through the maximum cascade time, $\tau_{\max}$ (maximized over $p$ for each $\zeta$).
% $\tau_{\max}$ peaks at $\zeta_c$; we denote this maximal timescale by $\tau_c\equiv \tau_{\max}(\zeta_c)$.
% This peak reflects critical slowing down at the crossover: below $\zeta_c$, activation grows continuously through local spreading, whereas above $\zeta_c$, activation is delayed by waiting for a supercritical nucleus and then accelerates abruptly.
% Therefore, $\zeta_c$ simultaneously marks (i) the minimum of $p_c^{(2)}$, (ii) the maximum timescale $\tau_c$, and (iii) the change in transition character at $p_c^{(2)}$ from continuous to discontinuous.

\subsection{\label{subsec:cascading_process}Cascade dynamics}
% The previous section demonstrated that the value of $\zeta$ controls the type of phase transition at $p_c^{(2)}$ and indicated the existence of  three distinct mechanisms.
% This section further clarifies the mechanisms responsible for the transitions, so it is instructive to examine the temporal evolution of the cascading activation process.
% Specifically, we select three values of $\zeta$: 3, 12 and 1000, near and far from criticality(see Fig.~\ref{fig:3}).
% Similar to Fig.~\ref{fig:1}(b,c), a fraction $p$ of nodes is randomly selected as the initial seed.
% We then follow the temporal evolution of the giant connected component $P_{\infty}(t)$, across iterations(time steps), as shown in Fig.~\ref{fig:3}(a)-(c).
% Furthermore, to clearly differentiate between the dynamical processes associated with the three distinct transition types, we track the fraction of nodes newly added to the giant connected component, $S_t$, in Fig.~\ref{fig:3}(d)-(f).
% We also monitor the branching factor, $\eta_t$, defined as the ratio between activation sizes at two successive time steps $t$ and $t-1$ in Fig.~\ref{fig:3}(g)-(i).

\begin{figure*}[htbp]
\centering
% 导入包含a-d子图的主图
\includegraphics[width=\textwidth]{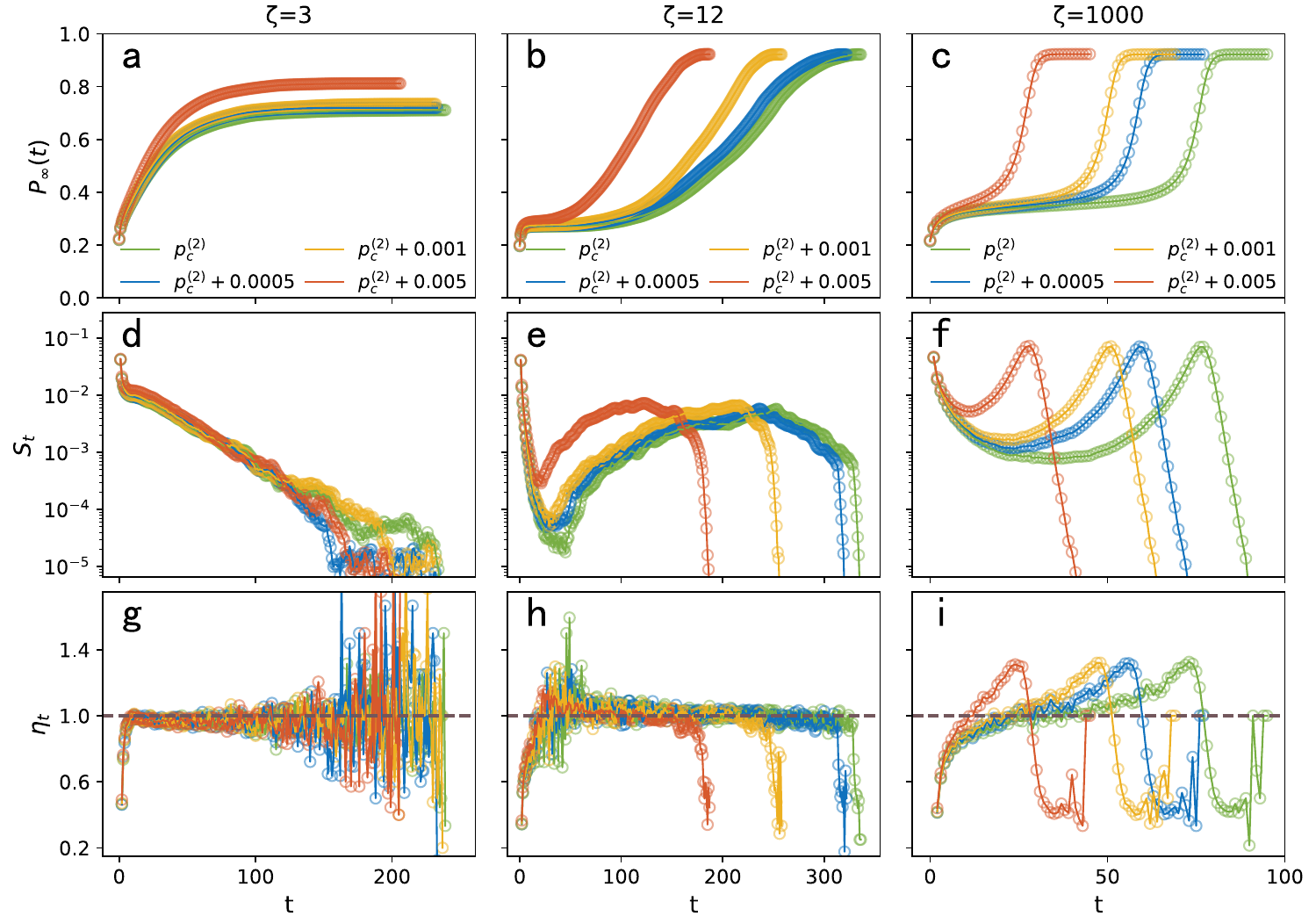}

% 主图图注
\caption{\textbf{Distinct cascading dynamics associated with the three routes to collective activation.}
  Bootstrap percolation dynamics are shown for three representative interaction ranges, $\zeta=3$, $12$, and $1000$ ($L=1000$), corresponding to continuous, purely first-order, and mixed-order phase transitions, respectively.
  Columns correspond to increasing interaction range:
  the left column (a,d,g) shows the strongly spatial regime $\zeta=3\ll L$;
  the middle column (b,e,h) shows the intermediate, nucleation-dominated regime $\zeta=12$;
  and the right column (c,f,i) corresponds to the effectively non-spatial, branching-dominated regime $\zeta=L=1000$.
  Different colors indicate different initial activation probabilities $p$ near criticality, with green curves highlighting the critical case $p=p_c^{(2)}$.
  (\textbf{a--c}) Time evolution of the relative size of the giant active cluster, $P_\infty(t)$.
  (\textbf{d--f}) Fraction of newly activated nodes added to the giant component at each iteration step, $S_t$.
  (\textbf{g--i}) Branching factor $\eta_t = S_t/S_{t-1}$ ($t>1$), characterizing the local amplification of activation during the cascade.
  In the strongly spatial regime ($\zeta=3$), activation proceeds via fractal growth: $S_t$ decays algebraically with time and the branching factor remains below unity, leading to gradual, continuous expansion.
  In the intermediate regime ($\zeta=12$), activation is governed by nucleation dynamics: after the formation of a critical nucleus, $S_t$ increases as the active region expands radially, until boundary effects suppress further growth.
  In contrast, in the non-spatial regime ($\zeta=L$), activation follows a branching process: $S_t$ remains microscopic over extended times with $\eta_t \approx 1$, before a sudden amplification ($\eta_t \gg 1$) triggers an avalanche-like cascade, producing the abrupt jump in $P_\infty(t)$.
  These sharply distinct temporal signatures provide direct dynamical evidence for the different mechanisms underlying the three types of phase transitions identified in Fig.~\ref{fig:2}(a).
  All networks have the same system size $N$ and mean degree $\langle k \rangle$ as in Figs.~\ref{fig:1}(b,c) and \ref{fig:2}.
}
\label{fig:3}  % 主图标签
\end{figure*}

The results above show that the interaction range $\zeta$ controls the nature of the transition at $p_c^{(2)}$.
To uncover the underlying mechanism, we examine the temporal evolution of the cascade.
Figure~\ref{fig:3} shows how the interaction range $\zeta$ selects distinct microscopic pathways to collective activation, as captured by the dynamics of the giant active component $P_{\infty}(t)$, the fraction of newly activated nodes $S_t$, and the branching factor $\eta_t$.

For small $\zeta$ (Fig.~\ref{fig:3}(a,d,g)), activation remains strongly local and does not develop into a self-sustained propagation mechanism.
The increment $S_t$ rapidly decays and remains small, while the branching factor stays below unity $\eta_t<1$, indicating that activation does not amplify.
As a result, growth proceeds through sparse, spatially heterogeneous activation, leading to a smooth increase of $P_{\infty}(t)$ even near $p_c^{(2)}$.
This behavior reflects a regime dominated by local, fractal-like spreading without collective amplification, consistent with a continuous transition.

At intermediate $\zeta \approx \zeta_c$ (Fig.~\ref{fig:3}(b,e,h)), the dynamics change qualitatively and become nucleation-driven.
After an initial transient, $P_{\infty}(t)$ grows linearly in time, signaling a propagating front with nearly constant velocity.
This transition is preceded by a peak in $S_t$ and a sustained regime with $\eta_t>1$, indicating that activation becomes self-amplifying once a critical nucleus is formed.
The subsequent expansion of this nucleus organizes the dynamics into a coherent front, producing a rapid macroscopic transition characteristic of a pure first-order transition.

For large $\zeta$ (Fig.~\ref{fig:3}(c,f,i)), the dynamics are no longer governed by spatial propagation but by a branching process.
The evolution of $P_{\infty}(t)$ exhibits a pronounced plateau followed by a sudden jump, indicating a clear separation of timescales.
During the plateau, $S_t$ remains microscopic and $\eta_t \approx 1$, placing the system in a marginally stable state where activation neither grows nor dies out on average.
Only when fluctuations drive $\eta_t>1$ does a rapid, system-wide cascade occur, leading to an abrupt increase in $P_{\infty}$.
This temporal pattern---a prolonged critical plateau followed by a sudden avalanche---is the hallmark of branching-driven dynamics.

These results establish a direct link between cascade dynamics and transition mechanisms:
Local, non-amplifying activation ($\eta_t<1$) leads to gradual, continuous growth;
self-sustained amplification via nucleation ($\eta_t>1$)
produces coherent front propagation and a first-order transition;
while marginal branching dynamics ($\eta_t \approx 1$) generate a critical plateau followed by a fluctuation-triggered avalanche.
In this sense, the interaction range $\zeta$ does not merely shift the transition point but determines how collective activation unfolds at the microscopic level.
Additional results for another activation threshold ($T=4$) are presented to support this classification (Supplementary Fig.~S3).

\subsection{\label{subsec:dynamical_scaling}Dynamical scaling}
\begin{figure}[htbp]
  \centering
  % 导入包含a-d子图的主图
  \includegraphics[width=0.5\textwidth]{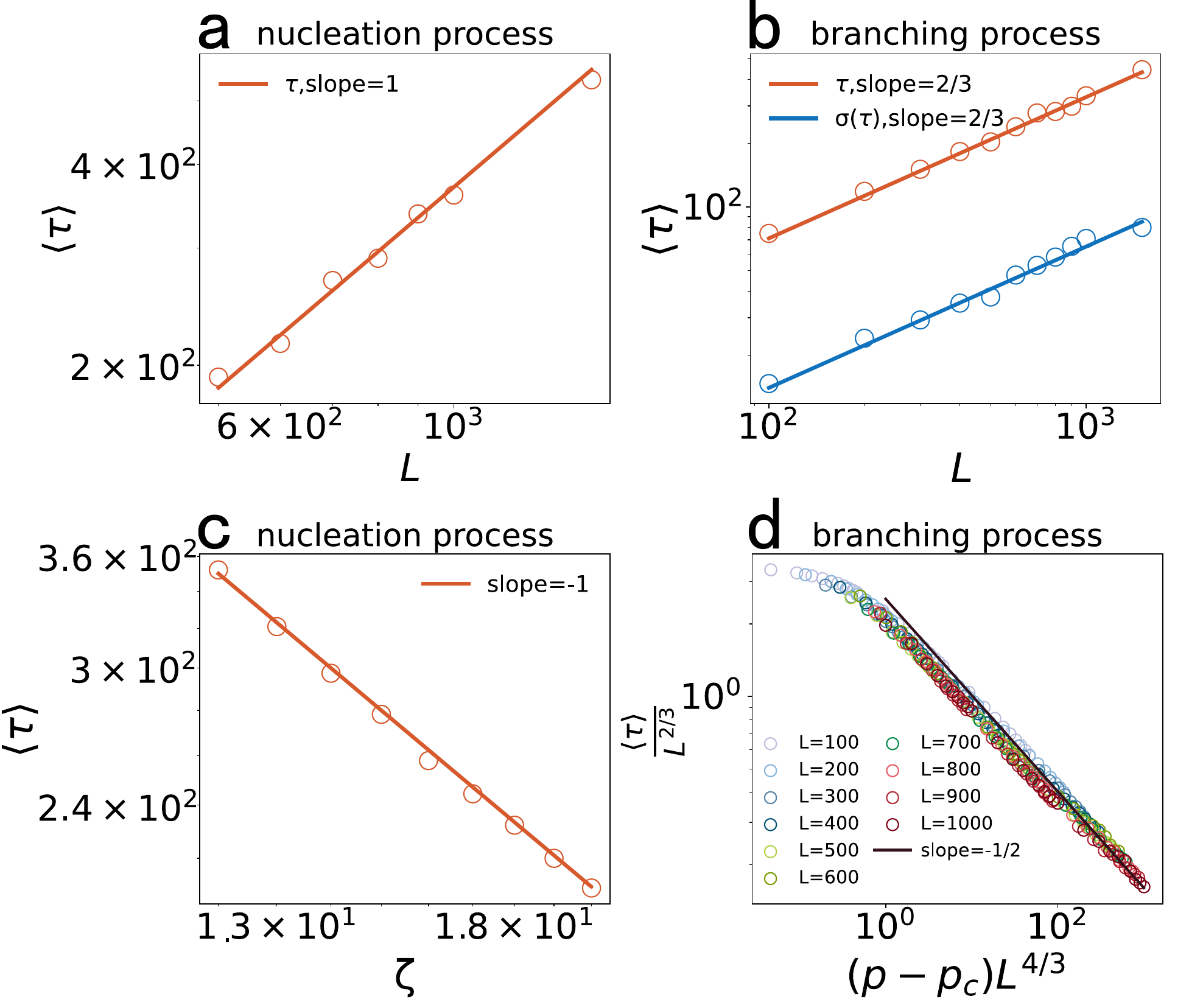}

  % 主图图注
  \caption{\textbf{Distinct scaling signatures of nucleation- and branching-driven cascades near the second transition point.}
    Panels (a,c) and (b,d) are grouped to contrast the scaling behavior associated with two fundamentally different cascading mechanisms, thereby providing dynamical evidence for the distinct phase transitions reported in the main text.
    \textbf{(a)} In the nucleation-dominated regime at intermediate interaction range ($\zeta=12$), the average cascade duration exhibits linear system-size scaling, $\langle \tau \rangle \sim L$, characteristic of a pure first-order transition driven by nucleation process.
    \textbf{(c)} In the same regime, the average cascade duration decreases inversely with the interaction range, $\langle \tau \rangle \sim \zeta^{-1}$ (system size $L=1000$), consistent with nucleation dynamics in which the cascade time scales $\tau \sim L / v$ with the radial propagation velocity $v$ increasing linearly with the interaction range $\zeta$.
    \textbf{(b)} In contrast, in the effectively non-spatial regime ($\zeta=L$), the cascade duration at the second critical point displays sublinear scaling, $\langle \tau \rangle \sim L^{2/3}$, accompanied by comparable fluctuations $\sigma(\tau)\sim L^{2/3}$, consistent with branching-dominated, mixed-order critical dynamics.
    \textbf{(d)} The same branching mechanism is further evidenced by the scaling of the cascade duration $\tau$ with the distance to criticality, $\Delta_p = p - p_c^{(2)}$.
    For panels (a) and (b), cascade durations are measured at the second critical point $p_c^{(2)}$ for different system sizes $L$.
    Data points represent averages over 50 independent realizations.
  }
  \label{fig:4}  % 主图标签
\end{figure}

Distinct microscopic cascade patterns indicate that the abrupt transition at $p_c^{(2)}$, marking the onset of collective activation, arise from different underlying mechanisms.
To distinguish between them, we focus on the cascade duration $\tau$ at $p_c^{(2)}$, which measures the intrinsic timescale of the activation process.
Unlike static observables, $\tau$ directly probes how activity propagates through the system and thus encodes the underlying microscopic mechanism.
Two distinct scaling behaviors emerge, providing a direct dynamical criterion to distinguish the underlying mechanisms of the transition.

In the nucleation-dominated regime (Fig.~\ref{fig:4}(a,c)), once a critical nucleus forms, the subsequent dynamics are governed by a compact activation front that propagates spontaneously.
The cascade proceeds via spatial propagation with a well-defined velocity $v$, so that the relaxation time is set by the traversal time across the system,
\begin{equation}
\langle \tau \rangle \sim L.
\end{equation}
This linear scaling~(Fig.~\ref{fig:4}(a)) indicates the dynamics are governed by front propagation.
Moreover, the front velocity increases with interaction range, $v \propto \zeta$ (see Supplementary Fig.~S4), leading to
\begin{equation}
\langle \tau \rangle \sim \zeta^{-1},
\end{equation}
as verified in Fig.~\ref{fig:4}(c).
Thus, the cascade is controlled by a single nucleation event followed by deterministic growth, consistent with classical first-order kinetics.

In contrast, in the branching-dominated regime (Fig.~\ref{fig:4}(b,d)), no stable activation front emerges.
Instead, the system enters a prolonged plateau stage, during which the cascade evolves through a critical branching process, as observed in Fig.~\ref{fig:3}(c).
In this regime, each activation event triggers on average one additional activation, placing the system at the edge of growth and extinction.
As a result, the cascade consists of marginally sustained avalanches with broadly distributed sizes and durations.
In the absence of a characteristic scale, the dynamics are governed by large fluctuations, leading to sublinear scaling,
\begin{equation}
\langle \tau \rangle \sim L^{2/3}, \qquad \sigma(\tau) \sim L^{2/3}.
\end{equation}
The identical scaling of the mean and its fluctuations reflects the underlying criticality.
Consistently, approaching the transition from above, the cascade duration diverges as
\begin{equation}
\tau \sim (p - p_c^{(2)})^{-1/2},
\end{equation}
(Fig.~\ref{fig:4}(d)), reflecting critical slowing down near the branching threshold.
This divergence mirrors second-order behavior despite the presence of a macroscopic discontinuity.
These results show that the system operates at a critical branching point embedded within a discontinuous transition:
the abrupt jump in the order parameter coexists with diverging spatiotemporal fluctuations, placing the transition in the mixed-order universality class~\cite{zhou2014simultaneous,Gao2024}.
This picture is supported by the finite-size correlation-length exponent (Supplementary Fig.~S5), consistent with hybrid universality classes of cascading failures~\cite{bonamassa2025hybrid}.

The contrast between $\langle \tau \rangle \sim \zeta^{-1}$ and $\langle \tau \rangle \sim L^{2/3}$ provides a direct dynamical signature of the underlying mechanisms:
ballistic front propagation driven by nucleation versus fluctuation-dominated spreading governed by critical branching.
This establishes that the two types of abrupt transitions are rooted in fundamentally different microscopic processes.

\subsection{\label{subsec:metastability}Metastability}
\begin{figure*}[htbp]
  \centering
  % 导入包含a-d子图的主图
  \includegraphics[width=\textwidth]{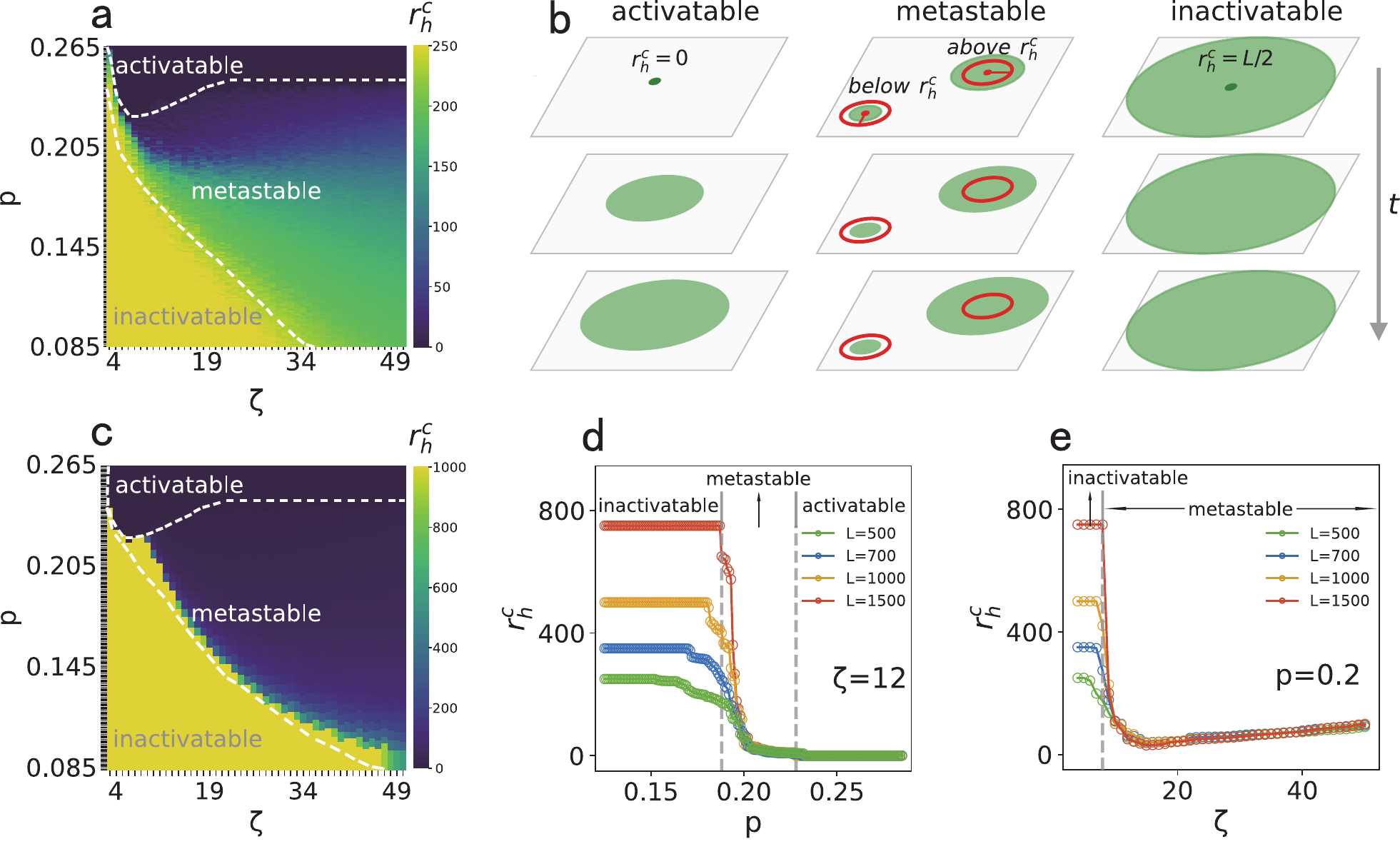}

  % 主图图注
  \caption{
    \textbf{Metastability and extreme susceptibility to local perturbations in spatial bootstrap percolation}.
    (a) Phase diagram in the ($\zeta$,$p$) plane,characterizing the system’s response to localized perturbations.
    The activatable phase (purple, $r^c_h = 0$) corresponds to spontaneous global activation from random seeding, while the inactivatable phase (yellow, $r^c_h \to L/2$) is stable against any finite localized perturbation, as illustrated schematically in panel (b).
    Between them lies a metastable region ($0 < r^c_h < L/2$),
    where global activation requires a localized seed exceeding a critical radius $r^c_h$.
    The color scale (green to blue) indicates the magnitude of $r^c_h$.
    System parameters: $N = L \times L = 500 \times 500$, $T=6$, and $\langle k \rangle=10$.
    (b) Schematic of activation propagation in the metastable regime triggered by a localized seed of radius $r_h$.
    Subcritical seeds ($r_h < r^c_h$) remain confined, while supercritical seeds ($r_h > r^c_h$) give rise to activation that propagates outward and expands radially over time $t$, leading to a system-wide cascade.
    (c) Phase diagram in the same ($\zeta$,$p$) plane as in (a), obtained from theory, showing the same three regimes: activatable, metastable, and inactivatable.
    (d) Critical radius $r^c_h$ as a function of the initial activation probability $p$ at fixed $\zeta=12$ for different system sizes $L$.
    The sharp drop of $r^c_h$ signals an abrupt transition across the metastability boundary, separating the inactivatable to metastable phase.
    (e) Critical radius $r^c_h$ as a function of the interaction range $\zeta$ at fixed $p=0.2$.
    The non-monotonic dependence indicates an abrupt transition from an inactivatable phase to a metastable phase, where the system becomes highly sensitive to parameter variations.
  }
  \label{fig:5}  % 主图标签
\end{figure*}

Bootstrap percolation has mostly been studied under random initial conditions, where global activation is driven by a finite density of initially active nodes. 
In this setting, the transition is characterized by the critical threshold $p_c^{(2)}$, above which a system-wide cascade emerges. 
In spatially embedded networks, however, the finite interaction range $\zeta$ introduces strong spatial correlations and qualitatively alters the activation dynamics. 
These correlations generate nucleation-driven cascades that are absent in the random-seeding picture.
This shift raises a central question: how does the system respond to localized perturbations? 
We find that even below $p_c^{(2)}$, where random seeding alone fails, a finite localized perturbation can still trigger a global cascade. 
Thus, in this regime the system is not truly stable but metastable.

To characterize this behavior, we probe the response to localized activation. 
Starting from the steady state reached under random initial conditions at a given $p$, we activate all nodes within a circular region of radius $r_h$ and monitor the evolution. 
Our numerical and theoretical results reveal three distinct regimes in the $(p,\zeta)$ parameter space~(Fig.~\ref{fig:5}(a,c)). 
In the activatable phase, global activation is achieved by random seeding alone, corresponding to a vanishing critical radius $r_h^c=0$. 
In contrast, the inactivatable phase is stable against any finite localized perturbation, with $r_h^c \to L/2$. 
Between these two limits lies a metastable regime, where global activation requires a localized seed larger than a finite critical radius $r_h^c$. 

Within the metastable regime, the response to localized perturbations exhibits a clear threshold. 
As illustrated in Fig.~\ref{fig:5}(b), subcritical seeds ($r_h < r_h^c$) remain spatially confined, whereas supercritical ones ($r_h > r_h^c$) trigger a self-sustained nucleation process that expands outward and eventually activates the entire system. 
This threshold defines a critical radius $r_h^c$ separating confined from propagating activation.
The critical radius in the metastable phase remains finite and independent of system size (Fig.~\ref{fig:5}(d,e)), indicating that it is an intrinsic property of the system.
Consequently, even microscopic localized perturbations can induce macroscopic cascades in arbitrarily large systems, revealing an intrinsic susceptibility of spatial bootstrap percolation.

The metastable phase is governed by the stability of the activation front within a boundary region surrounding a localized seed. 
Within this region, activation depends on two factors: local spatial coupling, set by $\zeta$, and the random active background, set by $p$. 
If the local activation level remains above the self-sustaining threshold set by $T$, and inactive nodes do not form spanning gaps that disconnect the active interior, boundary nodes activate successively and the front propagates outward. 
If either condition fails, sustained activation cannot be maintained, and the front stalls. 
As a result, the metastable state is jointly determined by $\zeta$, $p$, and $T$, which together set the critical seed size $r_h^c$. 
A detailed analysis is provided in Appendix~\ref{sec:theory_rc}.

To further resolve the phase structure, we then examine how the critical radius $r_h^c$ varies with the control parameters. 
As shown in Fig.~\ref{fig:5}(d,e), for fixed $\zeta$ or $p$, $r_h^c$ exhibits a sharp jump at the boundary between the inactivatable phase and the metastable regime. 
For small system sizes, this transition is smoothed by finite-size effects, but it becomes sharper as the system size increases, indicating that it is intrinsic.
This discontinuity reveals an abrupt transition: even a small change in the control parameters can qualitatively alter the system’s response to localized perturbations. 
Across this transition, the system switches from being robust against any finite perturbation to being activatable by a localized seed of finite size. 
Accordingly, the activation dynamics change discontinuously rather than gradually, reflecting strong sensitivity to control parameters.

These results establish metastability as an intrinsic property of spatial bootstrap percolation. 
Global activation is controlled by a finite, size-independent critical nucleus, allowing microscopic localized perturbations to trigger macroscopic cascades. 
Moreover, the transition from the inactivatable phase to the metastable regime is abrupt, underscoring strong sensitivity to external perturbations.

\section{\label{sec:discussion}Discussion}
In this work, we show that the interaction range fundamentally reshapes threshold-driven activation by coupling microscopic mechanisms to macroscopic critical behavior. 
By tuning $\zeta$, the system exhibits three distinct regimes: continuous, first-order, and mixed-order transitions, showing that the nature of the phase transition is not fixed but selected by the underlying activation dynamics. 
A central result is the identification of two qualitatively different activation mechanisms at the transition: nucleation-driven front propagation and critical branching. 
These mechanisms give rise to distinct dynamical scaling behaviors and account for the observed differences in both the order parameter and cascade dynamics. 
In particular, we show that first-order and mixed-order transitions at $p_c^{(2)}$, although both discontinuous at the macroscopic level, arise from fundamentally different microscopic mechanisms, as reflected in their scaling properties and temporal evolution.
These results provide a general framework for understanding different types of phase transitions in a broad class of systems. 

Another key finding is the emergence of a metastable phase near the first-order transition. 
In this regime, the system appears stable under random seeding, yet can be destabilized by a finite localized perturbation. 
The critical nucleus required to trigger global activation remains finite and does not scale with system size, implying that microscopic perturbations can induce macroscopic cascades even in arbitrarily large systems. 
This reveals an intrinsic, scale-independent susceptibility of spatial threshold-driven systems and provides direct evidence for metastability in spatial networks.
Moreover, the transition from the inactivatable phase to the metastable regime is abrupt, indicating that even small changes in control parameters can qualitatively alter the system’s response to external perturbations.

Our results stand in contrast to the traditional understanding in non-spatial networks, such as Erd\H{o}s--R\'enyi networks, where bootstrap percolation typically displays a single hybrid (mixed-order) transition governed by mean-field criticality \cite{dorogovtsev2006k,baxter2011heterogeneous}. 
They also differ from previous studies of bootstrap percolation on spatial networks~\cite{gao2015bootstrap}, where spatial effects are introduced through heterogeneous link distributions rather than a well-defined interaction range. 
In such models, the effective spatial coupling is not explicitly controlled, obscuring the emergence of distinct transition regimes.
In contrast, by directly tuning the interaction range $\zeta$, our framework provides precise control of spatial coupling, revealing how it selects both the order of the transition and the dominant activation pathways. 
This establishes interaction range as a key control parameter governing both cascade dynamics and transition mechanisms in spatial systems.

%More broadly, our results suggest that interaction range acts as a unifying control parameter for activation dynamics in spatial systems. 
Similar crossovers between nucleation-controlled and branching-driven behavior have been observed not only in spatial $k$-core percolation~\cite{xue2024nucleation}, but also in a broader class of cascading systems, including interdependent percolation~\cite{li2012cascading} and related failure processes~\cite{gross2024microscopic}. 
The recurrence of these phenomena across different models suggests that they are not system-specific, but instead reflect a shared spatial constraint.
In this context, interaction range emerges as a unifying control parameter, organizing similar cascade behaviors across systems and shaping their underlying mechanisms.
More broadly, this suggests that cascade dynamics may be organized into universality-like classes based on interaction range, analogous to critical phenomena.

\section*{Acknowledgements}
This work was supported by the Guangdong Basic and Applied Basic Research Foundation (Grant No.~2024A1515012692).

%\section*{\label{sec:odod}Methods}
%\subsection{\label{subsec:net_model}The $\zeta$-model}
\appendix
\section{\label{sec:net_model}The $\zeta$-model}
We construct a spatially embedded undirected network consisting of $N = L^{2}$ nodes arranged on a two-dimensional periodic square lattice.
Each node corresponds to a lattice site with integer coordinates $(x,y)$, and the network is characterized by a prescribed average degree $\langle k \rangle$.
The spatial coupling is introduced through the characteristic link length~$\zeta$, which controls the distribution of Euclidean distances between connected node pairs.
To generate the network, we first draw $N\langle k\rangle/2$ independent random link lengths $l$ from the exponential distribution $ P(l) \sim e^{-l/\zeta} $ and restrict all sampled lengths to $l \leq L/2$ due to periodic boundary conditions.
For each desired link length~$l$, we randomly select a source node $(x_{0},y_{0})$ and attempt to identify a destination node located at Euclidean distance~$l$ from it.
Since the lattice is discrete, integer solutions to $l = \sqrt{\Delta x^{2} + \Delta y^{2}}$ may be absent for some values of~$l$.
In such cases, we adopt a nearest-distance matching procedure: among all lattice sites, we select the node whose distance from $(x_{0},y_{0})$ deviates the least from the target length~$l$, and connect these two nodes.
We enforce simple-graph constraints by forbidding self-loops and multiple edges between the same pair of nodes.
The above sampling and matching process is repeated until $N\langle k\rangle/2$ distinct edges have been successfully generated.
The resulting spatial network reproduces the prescribed exponential link-length distribution and tunes long-range connectivity through~$\zeta$.

%\subsection{Gyration Radius}
%\label{subsec:measure_size}
\section{\label{sec:measure_size}Gyration Radius}
To quantify spatial growth during bootstrap cascades, we track the largest active connected component at each update step.
For each realization with initial activation probability $p$ and threshold $T$, dynamics run until no further activations occur.
At cascade step $t$, let $C_t$ be the largest connected component (LCC) of active nodes.
Because each node has a fixed coordinate on the two-dimensional periodic lattice, $C_t$ can be mapped directly to Euclidean space and characterized geometrically.

We measure its spatial extent by the radius of gyration,
\[
R_g^2(t)=\frac{1}{|C_t|}\sum_{i\in C_t} d_{\mathrm{PBC}}^2\!\left(\mathbf{r}_i,\mathbf{r}_{\mathrm{cm}}(t)\right),
\]
where $\mathbf{r}_i=(x_i,y_i)$ is the position of node $i$, $\mathbf{r}_{\mathrm{cm}}(t)$ is the center of mass of $C_t$ computed with periodic wrapping, and $d_{\mathrm{PBC}}$ denotes Euclidean distance under periodic boundary conditions.
The resulting time series $R_g(t)$ provides a robust proxy for the effective size of the active core and enables a direct comparison of growth modes (continuous spreading, nucleation-driven expansion, and avalanche-like propagation) across different interaction ranges $\zeta$.

%\subsection{Mean-field prediction}%of $p_c^{(2)}$
%\label{subsec:theoretical_threshold}
\section{\label{sec:theoretical_threshold}Gyration Radius}
To benchmark the second transition point $p_c^{(2)}$, we analyze bootstrap percolation on an Erdős--Rényi (ER) graph, which is the mean-field limit of the spatial model at large interaction range $\zeta$.
In an ER graph with mean degree $\langle k\rangle$, each node is initially active with probability $p$, and an inactive node activates once at least $T$ neighbors are active.
Let $q$ be the probability that a randomly followed edge reaches a node that is eventually active.
Under the locally tree-like approximation, $q$ obeys
\begin{equation}
q = p + (1-p)\sum_{m=T}^{\infty}\frac{(\langle k\rangle q)^m}{m!}e^{-\langle k\rangle q}.
\label{eq:self_consistency_q}
\end{equation}
A macroscopic active phase appears when a non-trivial stable fixed point emerges.
Accordingly, $p_c^{(2)}$ is determined by the saddle-node conditions
\begin{equation}
f(q)=q,\qquad \frac{df}{dq}=1,
\label{eq:saddle_node_condition}
\end{equation}
where $f(q)$ is the right-hand side of Eq.~(\ref{eq:self_consistency_q})~\cite{baxter2011heterogeneous,janson2012bootstrap}.

For $T=6$ and $\langle k\rangle=10$, solving Eqs.~(\ref{eq:self_consistency_q}) and (\ref{eq:saddle_node_condition}) yields
\begin{equation}
p_c^{(2)} \approx 0.242083.
\end{equation}
In the spatial model, increasing $\zeta$ progressively suppresses spatial correlations.
In the large-$\zeta$ regime ($\zeta=1000$ in our simulations), the system converges effectively to the ER ensemble with the same $\langle k\rangle$.
Under identical parameters, simulations give
\begin{equation}
p_c^{(2)}(\zeta=1000)=0.242127,
\end{equation}
averaged over more than 50 realizations.
The quantitative agreement between theory and simulation confirms that the large-$\zeta$ regime is governed by mean-field behavior, and therefore provides a controlled reference for identifying deviations induced by spatial constraints at finite $\zeta$.

%下面对应$r_h$以及$r_h^c$
%\subsection{Critical nucleation radius}% $r_h^c$
%\label{subsec:theory_rc}
\section{\label{sec:theory_rc}Critical nucleation radius}
To characterize nucleation onset, we consider a localized active seed embedded in a metastable background.
The system first relaxes to a stationary state with active density $p_\infty(\zeta,p)$.
We then introduce a fully active disk of radius $r_h$ on this background and examine whether it remains confined or drives a propagating cascade.

Because interactions have finite range $\zeta$, the seed induces a spatial activation gradient \cite{berezin2015localized}.
Nodes outside the seed are influenced only through overlap between their interaction neighborhood and the active disk, so activation is strongest near the boundary and decays with distance.
For a node at distance $\rho$ from the seed boundary, the effective fraction of active neighbors is
\begin{equation}
p_{\mathrm{eff}}(\rho)
=
p_\infty
+
(1-p_\infty)\frac{I(r_h,\zeta,r_h+\rho)}{\pi \zeta^2},
\end{equation}
where $I(r_h,\zeta,r_h+\rho)$ is the geometric overlap between the seed and the interaction disk.
Since this overlap decreases monotonically with $\rho$, the activation field decays smoothly away from the seed.
Within a mean-field approximation, the corresponding local activation level is
\begin{equation}
P_{\mathrm{act}}(\rho)
=
\sum_{j=T}^{\langle k \rangle}
\binom{\langle k \rangle}{j}
\left[p_{\mathrm{eff}}(\rho)\right]^j
\left[1-p_{\mathrm{eff}}(\rho)\right]^{\langle k \rangle-j}.
\label{eq:Pact}
\end{equation}
which quantifies the propensity of nodes at distance $\rho$ to activate.

This profile defines an annular region around the seed where activation is locally favored.
For activation to spread outward from the seed, nodes in this annulus must activate in a self-sustained manner, i.e., they collectively satisfy the bootstrap condition.
The resulting activated annulus, together with the seed, forms an enlarged active core that can trigger activation in the next outer layer.
For this reason, the relevant local criterion is not mere geometric connectivity but the ability to sustain activation, which we characterize by the cascade threshold $p_c^{(2)}(\zeta)$.
Accordingly, we define a characteristic distance $\rho_c$ through
\begin{equation}
P_{\mathrm{act}}(\rho_c)=p_c^{(2)}(\zeta),
\label{eq:rhoc}
\end{equation}
which marks the distance beyond which activation can no longer sustain outward propagation.
Hence, $0<\rho<\rho_c$ is the annulus that can be incorporated into the growing active core.

To characterize the overall activation level within this annulus, we introduce the coarse-grained quantity
\begin{equation}
P_{\mathrm{ann}}
=
\frac{1}{\rho_c}\int_0^{\rho_c}P_{\mathrm{act}}(\rho)\,\mathrm{d}\rho.
\label{eq:Pann}
\end{equation}
This quantity represents an effective activation probability for the annulus, replacing the spatial gradient by a uniform activation level.
However, a sufficiently high activation level alone does not guarantee nucleation.
Even within this annulus, some nodes remain inactive and may form connected gaps.
If such gaps span the annulus, they disconnect the active interior from the exterior and block further activation.
If they remain shorter than the annulus width, the active nodes form a connected backbone and the front can continue to advance.

To capture this competition, we introduce a geometric scale $\xi_{\mathrm{inactive}}$, the typical size of connected inactive regions in the annulus.
This quantity is measured from connected components of inactive nodes and directly quantifies their ability to interrupt the front.
The nucleation condition is therefore
\begin{equation}
\xi_{\mathrm{inactive}} < \rho_c.
\end{equation}
The critical seed radius $r_h^c$ is determined by the marginal condition
\begin{equation}
\xi_{\mathrm{inactive}}(r_h^c,p_\infty,\zeta)=\rho_c(r_h^c,p_\infty,\zeta),
\end{equation}
which separates confined and propagating regimes.
In this framework, nucleation occurs when a localized seed generates an annular region that is both self-sustaining and connected.
Increasing the seed size broadens this region and facilitates outward propagation, whereas large inactive gaps can disrupt it and arrest the front.
The critical radius $r_h^c$ marks the onset of sustained front propagation.

\section*{Author contributions}
All authors have contributed equally to this article.

\section*{Competing interests}
The authors declare no competing interests.

\section*{Data \& Materials Availability}
%All data supporting the findings of this study are publicly available on xxx.
%The code required to reproduce the results of this study is also available on GitHub at \url{https://github.com/JialuZhang01/SpatialBootstrapPercolation}.
Both the data supporting the findings and the code required to reproduce the results of this study are publicly available on GitHub at \url{https://github.com/JialuZhang01/SpatialBootstrapPercolation}.
%Mendeley Data (\url{}), with detailed descriptions and usage notes provided in the Supplementary Materials.
%Zenodo (DOI: ; \url{}) 

\bibliography{reference}

@article{xue2024nucleation,
  title={Nucleation phenomena and extreme vulnerability of spatial k-core systems},
  author={Xue, Leyang and Gao, Shengling and Gallos, Lazaros K and Levy, Orr and Gross, Bnaya and Di, Zengru and Havlin, Shlomo},
  journal={Nature Communications},
  volume={15},
  number={1},
  pages={5850},
  year={2024},
  publisher={Nature Publishing Group UK London}
}

@article{saberi2015recent,
  title={Recent advances in percolation theory and its applications},
  author={Saberi, Abbas Ali},
  journal={Physics Reports},
  volume={578},
  pages={1--32},
  year={2015},
  publisher={Elsevier}
}

@article{gao2015bootstrap,
  title={Bootstrap percolation on spatial networks},
  author={Gao, Jian and Zhou, Tao and Hu, Yanqing},
  journal={Scientific Reports},
  volume={5},
  number={1},
  pages={14662},
  year={2015},
  publisher={Nature Publishing Group UK London}
}

@article{daqing2014spatial,
  title={Spatial correlation analysis of cascading failures: congestions and blackouts},
  author={Daqing, Li and Yinan, Jiang and Rui, Kang and Havlin, Shlomo},
  journal={Scientific Reports},
  volume={4},
  number={1},
  pages={5381},
  year={2014},
  publisher={Nature Publishing Group UK London}
}

@article{zhao2016spatio,
  title={Spatio-temporal propagation of cascading overload failures in spatially embedded networks},
  author={Zhao, Jichang and Li, Daqing and Sanhedrai, Hillel and Cohen, Reuven and Havlin, Shlomo},
  journal={Nature Communications},
  volume={7},
  number={1},
  pages={10094},
  year={2016},
  publisher={Nature Publishing Group UK London}
}

@article{holroyd2003sharp,
  title={Sharp metastability threshold for two-dimensional bootstrap percolation},
  author={Holroyd, Alexander E},
  journal={Probability Theory and Related Fields},
  volume={125},
  number={2},
  pages={195--224},
  year={2003},
  publisher={Springer}
}

@article{zhou2014simultaneous,
  title={Simultaneous first-and second-order percolation transitions in interdependent networks},
  author={Zhou, Dong and Bashan, Amir and Cohen, Reuven and Berezin, Yehiel and Shnerb, Nadav and Havlin, Shlomo},
  journal={Physical Review E},
  volume={90},
  number={1},
  pages={012803},
  year={2014},
  publisher={APS}
}

@article{baxter2010bootstrap,
  title={Bootstrap percolation on complex networks},
  author={Baxter, Gareth J and Dorogovtsev, Sergey N and Goltsev, Alexander V and Mendes, Jos{\'e} FF},
  journal={Physical Review E},
  volume={82},
  number={1},
  pages={011103},
  year={2010},
  publisher={APS}
}

@article{chalupa1979bootstrap,
  title={Bootstrap percolation on a Bethe lattice},
  author={Chalupa, John and Leath, Paul L and Reich, Gary R},
  journal={Journal of Physics C: Solid State Physics},
  volume={12},
  number={1},
  pages={L31},
  year={1979},
  publisher={IOP Publishing}
}

@article{centola2010spread,
  title={The spread of behavior in an online social network experiment},
  author={Centola, Damon},
  journal={Science},
  volume={329},
  number={5996},
  pages={1194--1197},
  year={2010},
  publisher={American Association for the Advancement of Science}
}

@article{goltsev2010stochastic,
  title={Stochastic cellular automata model of neural networks},
  author={Goltsev, AV and De Abreu, FV and Dorogovtsev, SN and Mendes, JFF},
  journal={Physical Review E},
  volume={81},
  number={6},
  pages={061921},
  year={2010},
  publisher={APS}
}

@article{wu2014multiple,
  title={Multiple hybrid phase transition: Bootstrap percolation on complex networks with communities},
  author={Wu, Chong and Ji, Shenggong and Zhang, Rui and Chen, Liujun and Chen, Jiawei and Li, Xiaobin and Hu, Yanqing},
  journal={Europhysics Letters},
  volume={107},
  number={4},
  pages={48001},
  year={2014},
  publisher={IOP Publishing}
}

@article{li2015percolation,
  title={Percolation transition in dynamical traffic network with evolving critical bottlenecks},
  author={Li, Daqing and Fu, Bowen and Wang, Yunpeng and Lu, Guangquan and Berezin, Yehiel and Stanley, H Eugene and Havlin, Shlomo},
  journal={Proceedings of the National Academy of Sciences},
  volume={112},
  number={3},
  pages={669--672},
  year={2015},
  publisher={National Academy of Sciences}
}

@inproceedings{danziger2013interdependent,
  title={Interdependent spatially embedded networks: dynamics at percolation threshold},
  author={Danziger, Michael M and Bashan, Amir and Berezin, Yehiel and Havlin, Shlomo},
  booktitle={2013 International Conference on Signal-Image Technology \& Internet-Based Systems},
  pages={619--625},
  year={2013},
  organization={IEEE}
}

@article{berezin2015localized,
  title={Localized attacks on spatially embedded networks with dependencies},
  author={Berezin, Yehiel and Bashan, Amir and Danziger, Michael M and Li, Daqing and Havlin, Shlomo},
  journal={Scientific Reports},
  volume={5},
  number={1},
  pages={8934},
  year={2015},
  publisher={Nature Publishing Group UK London}
}

@article{li2019vulnerability,
  title={Vulnerability analysis and critical area identification of public transport system: A case of high-speed rail and air transport coupling system in China},
  author={Li, Tao and Rong, Lili and Yan, Kesheng},
  journal={Transportation Research Part A: Policy and Practice},
  volume={127},
  pages={55--70},
  year={2019},
  publisher={Elsevier}
}

@article{zhang2025delayed,
  title={Delayed threshold and spatial diffusion in k-core percolation induced by long-range connectivity},
  author={Zhang, Xin-Ya and Yao, Yi and Han, Zhiyu and Yan, Gang},
  journal={Communications Physics},
  volume={8},
  number={1},
  pages={238},
  year={2025},
  publisher={Nature Publishing Group UK London}
}

@article{pang2023geometric,
  title={Geometric constraints on human brain function},
  author={Pang, James C and Aquino, Kevin M and Oldehinkel, Marianne and Robinson, Peter A and Fulcher, Ben D and Breakspear, Michael and Fornito, Alex},
  journal={Nature},
  volume={618},
  number={7965},
  pages={566--574},
  year={2023},
  publisher={Nature Publishing Group UK London}
}

@article{zhang2024geometric,
  title={Geometric scaling law in real neuronal networks},
  author={Zhang, Xin-Ya and Moore, Jack Murdoch and Ru, Xiaolei and Yan, Gang},
  journal={Physical Review Letters},
  volume={133},
  number={13},
  pages={138401},
  year={2024},
  publisher={APS}
}

@article{roberts2016contribution,
  title={The contribution of geometry to the human connectome},
  author={Roberts, James A and Perry, Alistair and Lord, Anton R and Roberts, Gloria and Mitchell, Philip B and Smith, Robert E and Calamante, Fernando and Breakspear, Michael},
  journal={Neuroimage},
  volume={124},
  pages={379--393},
  year={2016},
  publisher={Elsevier}
}

@article{firestone2011importance,
  title={The importance of location in contact networks: Describing early epidemic spread using spatial social network analysis},
  author={Firestone, Simon M and Ward, Michael P and Christley, Robert M and Dhand, Navneet K},
  journal={Preventive Veterinary Medicine},
  volume={102},
  number={3},
  pages={185--195},
  year={2011},
  publisher={Elsevier}
}

@article{zhang2016modeling,
  title={Modeling spatial contacts for epidemic prediction in a large-scale artificial city},
  author={Zhang, Mingxin and Verbraeck, Alexander and Meng, Rongqing and Chen, Bin and Qiu, Xiaogang},
  journal={Journal of Artificial Societies and Social Simulation},
  volume={19},
  number={4},
  year={2016},
  publisher={JASSS}
}

@article{danziger2016effect,
  title={The effect of spatiality on multiplex networks},
  author={Danziger, Michael M and Shekhtman, Louis M and Berezin, Yehiel and Havlin, Shlomo},
  journal={Europhysics Letters},
  volume={115},
  number={3},
  pages={36002},
  year={2016},
  publisher={IOP Publishing}
}

@article{ben1994generic,
  title={Generic modelling of cooperative growth patterns in bacterial colonies},
  author={Ben-Jacob, Eshel and Schochet, Ofer and Tenenbaum, Adam and Cohen, Inon and Czirok, Andras and Vicsek, Tamas},
  journal={Nature},
  volume={368},
  number={6466},
  pages={46--49},
  year={1994},
  publisher={Nature Publishing Group UK London}
}

@article{amini2010bootstrap,
  title={Bootstrap percolation in living neural networks},
  author={Amini, Hamed},
  journal={Journal of Statistical Physics},
  volume={141},
  number={3},
  pages={459--475},
  year={2010},
  publisher={Springer}
}

@article{breskin2006percolation,
  title={Percolation in living neural networks},
  author={Breskin, Ilan and Soriano, Jordi and Moses, Elisha and Tlusty, Tsvi},
  journal={Physical Review Letters},
  volume={97},
  number={18},
  pages={188102},
  year={2006},
  publisher={APS}
}

@article{grenfell2001travelling,
  title={Travelling waves and spatial hierarchies in measles epidemics},
  author={Grenfell, Bryan T and Bj{\o}rnstad, Ottar N and Kappey, Jens},
  journal={Nature},
  volume={414},
  number={6865},
  pages={716--723},
  year={2001},
  publisher={Nature Publishing Group UK London}
}

@article{hallatschek2014acceleration,
  title={Acceleration of evolutionary spread by long-range dispersal},
  author={Hallatschek, Oskar and Fisher, Daniel S},
  journal={Proceedings of the National Academy of Sciences},
  volume={111},
  number={46},
  pages={E4911--E4919},
  year={2014},
  publisher={National Academy of Sciences}
}

@article{baxter2011heterogeneous,
  title={Heterogeneous k-core versus bootstrap percolation on complex networks},
  author={Baxter, Gareth J and Dorogovtsev, Sergey N and Goltsev, Alexander V and Mendes, Jos{\'e} FF},
  journal={Physical Review E},
  volume={83},
  number={5},
  pages={051134},
  year={2011},
  publisher={APS}
}

@article{janson2012bootstrap,
  title={Bootstrap percolation on the random graph $G_{n,p}$},
  author={Janson, Svante and {\L}uczak, Tomasz and Turova, Tatyana and Vallier, Thomas},
  journal={Annals of Applied Probability},
  volume={22},
  number={5},
  pages={1989--2047},
  year={2012}
}

@article{dorogovtsev2006k,
  title={K-core organization of complex networks},
  author={Dorogovtsev, Sergey N and Goltsev, Alexander V and Mendes, Jose Ferreira F},
  journal={Physical Review Letters},
  volume={96},
  number={4},
  pages={040601},
  year={2006},
  publisher={APS}
}

@article{fujikawa1989fractal,
  title={Fractal growth of Bacillus subtilis on agar plates},
  author={Fujikawa, Hiroshi and Matsushita, Mitsugu},
  journal={Journal of the Physical Society of Japan},
  volume={58},
  number={11},
  pages={3875--3878},
  year={1989},
  publisher={The Physical Society of Japan}
}

@article{niemeyer1984fractal,
  title={Fractal dimension of dielectric breakdown},
  author={Niemeyer, Lucian and Pietronero, Luciano and Wiesmann, Hans J},
  journal={Physical Review Letters},
  volume={52},
  number={12},
  pages={1033},
  year={1984},
  publisher={APS}
}

@article{lechleiter1991spiral,
  title={Spiral calcium wave propagation and annihilation in Xenopus laevis oocytes},
  author={Lechleiter, James and Girard, Steven and Peralta, Ernest and Clapham, David},
  journal={Science},
  volume={252},
  number={5002},
  pages={123--126},
  year={1991},
  publisher={American Association for the Advancement of Science}
}

@article{brockmann2013hidden,
  title={The hidden geometry of complex, network-driven contagion phenomena},
  author={Brockmann, Dirk and Helbing, Dirk},
  journal={Science},
  volume={342},
  number={6164},
  pages={1337--1342},
  year={2013},
  publisher={American Association for the Advancement of Science}
}

@article{bak1987soc,
  title={Self-Organized Criticality: An Explanation of 1/f Noise},
  author={Bak, Per and Tang, Chao and Wiesenfeld, Kurt},
  journal={Physical Review Letters},
  volume={59},
  number={4},
  pages={381--384},
  year={1987},
  publisher={APS}
}

@article{olami1992ofc,
  title={Self-organized criticality in a continuous, nonconservative cellular automaton modeling earthquakes},
  author={Olami, Zeev and Feder, H J S and Christensen, Kim},
  journal={Physical Review Letters},
  volume={68},
  number={8},
  pages={1244--1247},
  year={1992},
  publisher={APS}
}

@article{sethna2001crackling,
  title={Crackling noise},
  author={Sethna, James P and Dahmen, Karin A and Myers, Christopher R},
  journal={Nature},
  volume={410},
  number={6825},
  pages={242--250},
  year={2001},
  publisher={Nature Publishing Group}
}

@book{debenedetti2020metastable,
  title={Metastable liquids: concepts and principles},
  author={Debenedetti, Pablo G},
  year={2020},
  publisher={Princeton University Press}
}

@article{Gao2024,
doi = {10.1088/1367-2630/ad1539},
url = {https://doi.org/10.1088/1367-2630/ad1539},
year = {2024},
month = {jan},
publisher = {IOP Publishing},
volume = {26},
number = {1},
pages = {013006},
author = {Gao, Shengling and Xue, Leyang and Gross, Bnaya and She, Zhikun and Li, Daqing and Havlin, Shlomo},
title = {Possible origin for the similar phase transitions in k-core and interdependent networks},
journal = {New Journal of Physics},
}

@article{bonamassa2025hybrid,
  title={Hybrid universality classes of systemic cascades},
  author={Bonamassa, Ivan and Gross, Bnaya and Kert{\'e}sz, Janos and Havlin, Shlomo},
  journal={Nature Communications},
  volume={16},
  number={1},
  pages={1415},
  year={2025},
  publisher={Nature Publishing Group UK London}
}

@article{li2012cascading,
  title={Cascading Failures in Interdependent Lattice Networks: The Critical Role of the Length of Dependency Links},
  author={Li, Wei and Bashan, Amir and Buldyrev, Sergey V and Stanley, H Eugene and Havlin, Shlomo},
  journal={Physical Review Letters},
  volume={108},
  number={22},
  pages={228702},
  year={2012},
  publisher={APS}
}

@article{gross2024microscopic,
  title={Microscopic intervention yields abrupt transition in interdependent ferromagnetic networks},
  author={Gross, Bnaya and Bonamassa, Ivan and Havlin, Shlomo},
  journal={Physical Review Letters},
  volume={132},
  number={22},
  pages={227401},
  year={2024},
  publisher={APS}
}
\end{document}